\documentclass[nofootinbib,11pt,preprintnumbers,floatfix]{revtex4-2}
\usepackage[utf8]{inputenc}

\usepackage[dvipsnames]{xcolor}
\definecolor{red}{rgb}{0.9, 0,0}
\definecolor{cerulean}{rgb}{0., 0.42,0.9}
\definecolor{navy}{rgb}{0.05, 0.05,0.8}

\usepackage[colorlinks]{hyperref}
\hypersetup{
	colorlinks = true,
	linkcolor = red,
	urlcolor  = blue,
	citecolor = blue,
	anchorcolor = blue
}

\usepackage{amsmath,amsfonts,amsthm,amssymb, bm, mathtools, latexsym} 
\usepackage{hyperref}
\usepackage{subcaption}
\usepackage{physics}
\usepackage{enumerate}
\usepackage[final]{showlabels}

\newcommand{\pd}{\partial}
\renewcommand{\o}{\omega}

\renewcommand{\d}{\delta}
\renewcommand{\t}{\theta}

\renewcommand{\l}{\lambda}

\newcommand{\D}{\Delta}

\NewDocumentCommand{\codeword}{v}{%
	\texttt{\textcolor{blue}{#1}}%
}

\usepackage{subfiles} 

\begin{document}
	
	\title{Stochastic Description of Near-Horizon Fluctuations in Rindler-AdS}
	
	\author{Yiwen Zhang and Kathryn M. Zurek}
	\affiliation{Walter Burke Institute for Theoretical Physics,\\
		California Institute of Technology, Pasadena, CA}

	\begin{abstract}
				We study quantum spacetime fluctuations near light-sheet horizons associated with a Rindler wedge in AdS spacetime, in the context of AdS/CFT. 
		In particular, we solve the vacuum Einstein equation near the light-sheet horizon, augmented with the Ansatz of a quantum source smeared out in a Planckian width along one of the light-cone directions. Such a source, whose physical interpretation is of gravitational shockwaves created by vacuum energy fluctuations, alters the Einstein equation to a stochastic partial differential equation taking the form of a Langevin equation. 
		By integrating fluctuations along the light sheet, we find an accumulated effect in the round-trip time of a photon to traverse the horizon of the Rindler wedge that depends on both the $d$-dimensional Newton constant $G_N^{(d)}$ and the AdS curvature $L$, in agreement with previous literature utilizing different methods.
	\end{abstract}
	
	\maketitle
	\newpage
	\tableofcontents
	\newpage
	
	%
	\section{Introduction}
	\label{sec:introduction}
	
\begin{figure}
		\centering
		\hspace*{-0.5in}
		\includegraphics[width=0.4\textwidth]{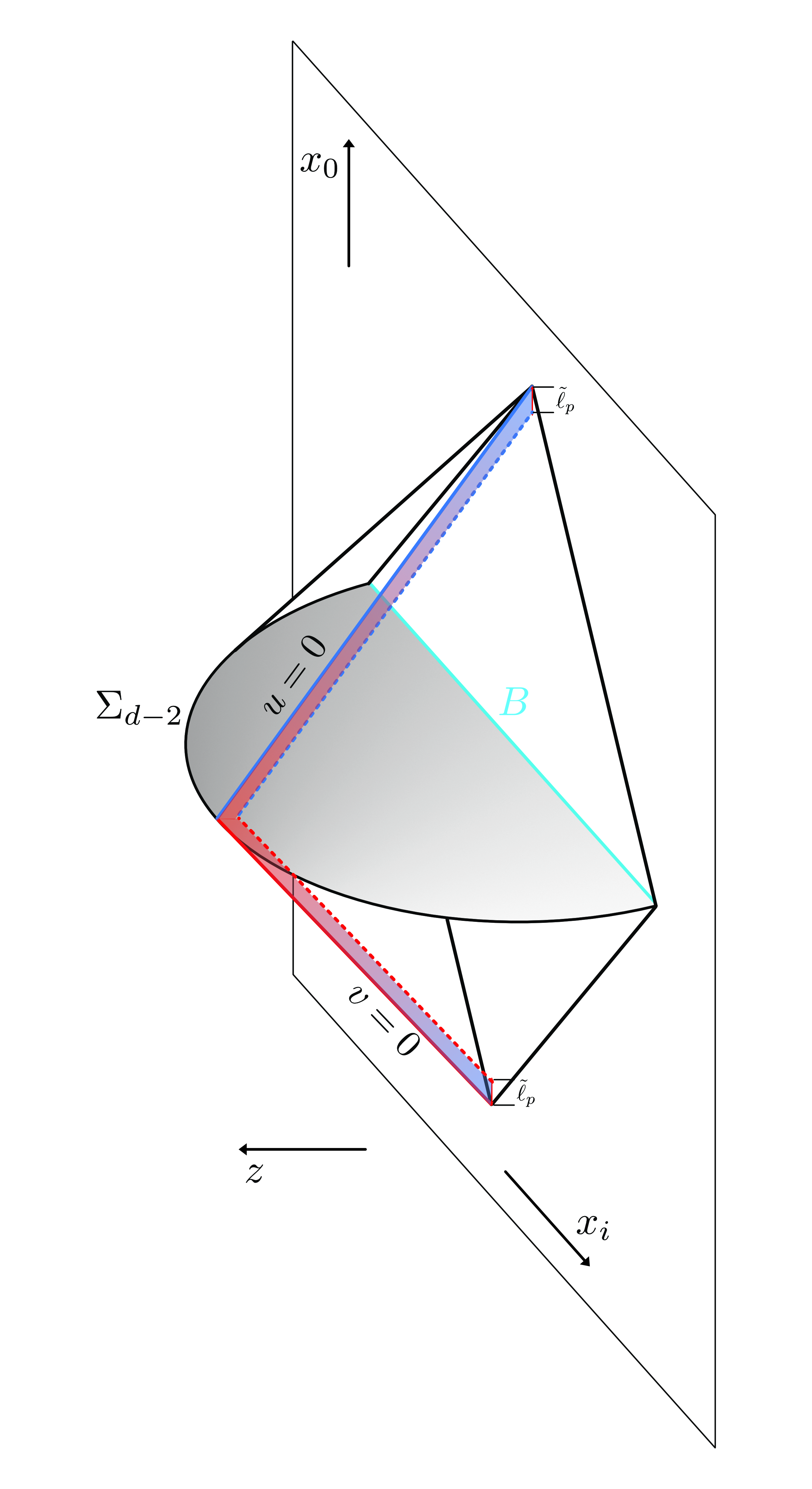} 
		\vspace*{-3mm}
		\caption{Depiction of the causal diamond in AdS space anchored at the boundary. The red  solid line traces out the light signal emitted from the boundary to a point in the bulk on the Ryu-Takayanagi surface labeled by $\Sigma_{d-2}$, while the blue solid line represents the light reflected from the point in the bulk and received at the boundary. The dashed lines represents the smearing of the light-sheet horizon. The red and blue shaded region represents quantum gravity induced fluctuations of the light trajectory.}\label{fig:causal_diamond} 
	\end{figure}

	The quantum mechanical description of gravity is one of the most elusive questions in physics. 
	An important tool towards understanding the ultimate theory of quantum gravity is the AdS/CFT correspondence.  In this paper, we aim to study the dynamics of gravity in the region of AdS spacetime near light sheets shown in Fig.~\ref{fig:causal_diamond}. In particular, we seek to understand how spacetime fluctuations alter the trajectory of a photon in the $d$-dimensional bulk.  Ref.~\cite{Verlinde:2019ade} found a fluctuation in the round-trip time, $T_{\rm r.t.}$, of a photon traveling from the AdS boundary to the Ryu-Takayanagi (RT) surface $\Sigma_{d-2}$ in the bulk having area $A(\Sigma_{d-2})$ and back to the boundary:
	\begin{equation} \label{eq:Uncert_T_rt_VZ2}
		\frac{\D T_{\rm r.t.}^{2}}{T_{\rm r.t.}^{2}} = \frac{2}{(d-2)} \sqrt{\frac{4 G_N^{(d)}}{A(\Sigma_{d-2})}}.
	\end{equation}

	The boundary of a causal diamond created by light sheets is defined by a Rindler horizon, which has a non-zero temperature and entropy, similar to a black hole event horizon. The calculation of Ref.~\cite{Verlinde:2019ade} (as well as Refs.~\cite{Verlinde:2019xfb,Banks:2021jwj}) operated through the analogue between the boundary of the Rindler wedge and a black hole horizon, utilizing techniques developed in {\em e.g.}~\cite{Casini:2011kv,Hung:2011nu}.  In AdS/CFT, the modular Hamiltonian $K$ and its fluctuations $\Delta K$ obey an area law similar to a black hole horizon \cite{Verlinde:2019ade,Perlmutter2014,Nakaguchi:2016zqi,DeBoer:2018kvc}
	\begin{equation}
		\label{eq:DeltaK}
		\langle K \rangle = \langle \Delta K^2 \rangle = \frac{A(\Sigma_{d-2})}{4 G_N^{(d)}}=S_{\rm ent.},
	\end{equation}
	where $S_{\rm ent.}$ is the entanglement entropy.	 Further, the metric, if restricting to only the part of the spacetime covered by the Rindler wedge shown in Fig.~\ref{fig:causal_diamond}, can be parameterized in terms of the {\em topological} black hole:
	\begin{equation} \label{eq:metric_top_BH}
		d s^2 = - f(\rho) d\tau^2 + \frac{d \rho^2}{f(\rho)} +  \frac{\rho^{2}}{L^{2}}	d \Sigma_{d-2}^{2} \qquad \mbox{ with} \qquad  f(\rho) = \frac{\rho^{2}}{L^{2}} - 1,
	\end{equation}
	where $L$ is the AdS radius, the radial coordinate $\rho$ ranges from $ L\le \rho  < \infty$.  Ref.~\cite{Zurek:2022xzl}, based on the calculations of Ref.~\cite{Verlinde:2019xfb,Verlinde:2019ade,Banks:2021jwj}, proposed a dictionary between the horizons of causal diamonds (in common spacetimes such as AdS and Minkowski) and black hole horizons.
	
	It has long been known that black hole horizons have a hydrodynamic description, known as the fluid-gravity correspondence~\cite{Damour:1979wya,Thorne_1986,Parikh:1997ma}.  The fluid-gravity correspondence was made more precise in the context of AdS/CFT, where the hydrodynamics of a strongly interacting fluid (\textit{e.g.,} quark-gluon plasma) on the asymptotic boundary of a lower-dimensional spacetime is described by gravitational dynamics on a black brane in the bulk of AdS~\cite{Policastro:2002se,Policastro:2002tn,Kovtun:2003wp,Kovtun:2005ev}.  These works inspired an extensive literature studying a hydrodynamic  effective description of gravity {\em e.g.}~\cite{Bhattacharyya:2007vjd,Rangamani:2009xk,deBoer:2015ija,Blake:2017ris,Nickel:2010pr,Crossley:2015tka}.  Further, Refs.~\cite{Bredberg:2010ky,Bredberg:2011jq} studied the dynamics of gravity in flat spacetime with a cut-off surface, showing that the Einstein Equation in vacuum reduces to a Navier-Stokes equation in one lower spacetime dimension.
	
	Here, we utilize an effective fluid description of gravity at the horizon of the Rindler Wedge in AdS shown in Fig.~\ref{fig:causal_diamond} to 	understand and re-cast the result Eq.~\eqref{eq:Uncert_T_rt_VZ2}. In particular, we will study the Einstein equation near the boundary defined by null sheets in Fig.~\ref{fig:causal_diamond}.  The hydrodynamic behavior of the metric becomes apparent when the vacuum Einstein equation in the near-horizon limit is augmented with an {\em Ansatz} that the Einstein equation has a quantum source:
	\begin{align}
		\begin{split} \label{eq:two-pt hh}
			\qty(\frac{d-2}{L^2} - \laplacian_{\perp}) \expval{h_{uu}(u,\vb{x}_\perp) h_{uu}(u',\vb{x}_\perp') } &= 8 \pi G_N^{(d)} \frac{\d(u - u')}{2\pi \tilde{\ell}_{p}} \d^{d-2}(\vb{x}_\perp - \vb{x}'_\perp), \\
			\qty(\frac{d-2}{L^2} - \laplacian_{\perp}) \expval{h_{vv}(v,\vb{x}_\perp) h_{vv}(v',\vb{x}_\perp') } &= 8 \pi G_N^{(d)} \frac{\d(v - v')}{2\pi\tilde{\ell}_{p}} \d^{d-2}(\vb{x}_\perp - \vb{x}'_\perp).
		\end{split}
	\end{align}
	Here, $u,~v$ are light-cone coordinates, while ${\bf x}_\perp$ are the $(d-2)$ remaining transverse spatial directions.  The left-hand side is derived from the vacuum Einstein Equation in AdS in the near-horizon limit, while the right-hand side is a quantum noise term, an {\em Ansatz} motivated by the membrane paradigm.    In particular, a gravitationally coupled ultra-local quantum noise term, $\delta^d(x-x')$, is reduced on one of the light-cone directions by smearing one of the light-cone delta functions with a Planckian width $\tilde \ell_p$ across a membrane (or black-brane) at the light-sheet horizon.  This smearing is depicted as a red/blue band in Fig.~\ref{fig:causal_diamond}. When we solve this equation to obtain the fluctuation in the photon round-trip traversal time, we will reproduce Eq.~\eqref{eq:Uncert_T_rt_VZ2}, provided that the width of the black-brane $\tilde \ell_p$ is the reduced Planck-scale, which we discuss below.  
	
	Note that the quantum source on the right-hand side of Eq.~\eqref{eq:two-pt hh} now appears like an energetic particle that creates a gravitational shockwave, as proposed by Dray-'t Hooft~\cite{DRAY1985173}.  Such shockwaves were recently shown in Ref.~\cite{Verlinde:2022hhs} to generate the modular relations in Eq.~\eqref{eq:DeltaK}, creating a self-consistent physical picture.  The quantum noise term in Eq.~\eqref{eq:two-pt hh} turns the Einstein equation into a Langevin-type equation 
	\begin{equation} \label{eq:langevin}
		\expval{\dot{X}(\tau) \dot{X}(\tau')} = \expval{F(\tau) F(\tau')},
	\end{equation} 
	where $\langle F(\tau) F(\tau') \rangle = 2 \mathcal{D} \delta(\tau-\tau')$ is a noise term with the diffusion coefficient $\mathcal{D}$ characterizing the scale of interaction, and we have integrated Eq.~\eqref{eq:two-pt hh} over the $(d-2)$ directions transverse to the lightcone coordinates.  Here $X(\tau)$ is a position variable identified with $X(\tau) = \int^{\tau} h_{\tau \tau}(\tau') d\tau'$, where $\tau=u(v)$ on the lower (upper) half of the causal diamond, and the two-point of $\eta(\tau)$ describes a stochastic noise that drives a random walk.  Consequently, the classical Einstein equation becomes a stochastic differential equation, where the {\em quantum uncertainty} in spacetime itself undergoes a random walk, with the correlations in the $(d-2)$ transverse directions given by the Green function of the transverse Laplacian.

	A ``smeared-out'' horizon is quite analogous to the notion of a stretched horizon, which is a time-like hypersurface Planckian separated from the true horizon first proposed by Damour~\cite{Damour:1979wya}. Later Refs.~\cite{Thorne_1986,Parikh:1997ma} showed that the properties of a black hole horizon can be mapped to those on the stretched horizon.  In the present context, we will smear out the horizon by a ``reduced Planck length'' previously identified in Refs.~\cite{Verlinde:2022hhs,Banks:2021jwj,Zurek:2022xzl}: 
	\begin{equation}\label{eq:ell_p tilde ansatz}
		\tilde \ell_p^2 \sim \frac{\ell_p^{d-2}}{L^{d-4}}.
	\end{equation}
	In four dimensions, the reduced Planck length corresponds to simply the Planck length,  $\tilde \ell_p \sim \ell_p \equiv \sqrt{8 \pi G_N}$.\footnote{In $d>4$, the Planck length is reduced by the IR scale $L$ to a scale smaller than $\ell_p$, suggesting to us that in $d>4$ there is actually no cumulative IR effect of the quantum fluctuations of spacetime.}  In a general number of dimensions, the length scale  in Eq.~\eqref{eq:ell_p tilde ansatz} was identified as the fundamental length scale of 't Hooft commutation relations in any number of dimensions~\cite{Verlinde:2022hhs}.  $\tilde \ell_p \sim L/\sqrt{S_{\rm ent.}}$ in Eq.~\eqref{eq:ell_p tilde ansatz} was also identified in Ref.~\cite{Banks:2021jwj} as the decoherence scale of nested causal diamonds, each of which have $S_{\rm ent}$ degrees-of-freedom.  We will find that positing a causal diamond with a stretched horizon of width given by Eq.~\eqref{eq:ell_p tilde ansatz} allows us to reproduce Eq.~\eqref{eq:Uncert_T_rt_VZ2}, the main result of Ref.~\cite{Verlinde:2019ade}.

	Lastly, we comment that while the square-root behavior of the variance $\D T_{\rm r.t.}^{2}$ in Eq.~\eqref{eq:Uncert_T_rt_VZ2} is perhaps somewhat perplexing from a scattering amplitude or na\"{i}ve EFT perspective, it is, however, characteristic of random walk behavior in hydrodynamics, where fluctuations take a typical form 
	\begin{equation} \label{eq:var Trt ansatz}
		\Delta T_{\rm r.t.}^{2} \sim \tilde \ell_p^2 {\cal N},
	\end{equation}
	where $\tilde \ell_p$ is the UV time scale of the hydrodynamic theory (normally associated to the diffusion coefficient, as discussed in Ref.~\cite{Zurek:2022xzl}), and ${\cal N} = L/ \tilde \ell_p$ is the number of steps in the random walk over the round-trip time.

	The outline of this paper is as follows. In Sec.~\ref{sec:background_geometry} we set the stage by reviewing the background geometry. In Sec.~\ref{sec:NH_metric_fluc}, we study the gravitational perturbations to the background geometry and show that the Einstein equation in near-horizon limit reduces to an equation relating metric fluctuations and gravitational shockwaves. In Sec.~\ref{sec:quant_noise}, we solve this equation with a source term derived from the t'Hooft commutation relations. Then we use the solution to calculate the uncertainty in photon round-trip time. Finally, in Sec.~\ref{sec:summery}, we discuss implications of our results and point to a few future directions. Throughout this paper, we will use $8 \pi G_{N}^{(d)} = \ell_{p}^{d-2}$ for the gravitational constant and Planck length in $d$ dimensions.

	%
	
	\section{Preliminaries: Background Geometry}
	\label{sec:background_geometry}
	
	As discussed in the introduction, we consider the geometric setup in Fig.~\ref{fig:causal_diamond}.
	A photon is emitted from the boundary into the bulk of $d$-dimensional AdS space, reflected by a ``mirror'' on the RT surface in the bulk, and finally received on the boundary. We briefly review three coordinates used throughout this paper, Poincar\'{e}, Eddington-Finkelstein (EF), and Kruskal–Szekeres.  The first (and most standard) will be useful for interpreting the results in terms of the observable time delay. The EF coordinates are closely related to topological-black hole coordinates introduced in Eq.~\eqref{eq:metric_top_BH}, and are useful as an intermediate step to derive equations of motion governing the dynamics of near-horizon metric fluctuations. Finally,  the Kruskal–Szekeres coordinates are the curved space analog of the light-cone metric. The light-cone metric is used extensively to study the effects of spacetime fluctuations of a causal diamond in Minkowski space in Refs.~\cite{Verlinde:2019xfb,Verlinde:2022hhs}, and a natural generalization to curved spacetime is provided by the Kruskal–Szekeres coordinates. We now proceed to briefly summarize these three coordinate systems, as useful for our discussion.
	
	\subsection{From Topological Black Hole to Poincar\'{e} Metric} \label{sub:poincare to Top BH}
	
	While the topological black hole metric described in Sec.~\ref{sec:introduction} is suitable to study the interior of the bulk causal diamond, there is also a causal diamond with spherical  symmetry on the boundary, such that the interferometer could also be viewed as being on the (suitably regularized) boundary. The Poincar\'{e} metric describes the near boundary region of the $d$-dimensional AdS space:
	\begin{equation} \label{eq:poincare metric}
		d s^2 = \frac{L^2}{z^2} (d z^2 - d x_0^2 + \delta_{a b} d x^a d x^b) \qfor a,b = 1, \dots, d-2.
	\end{equation}
	A causal diamond in AdS is illustrated in Fig.~\ref{fig:causal_diamond}, in which the blue line denoted by $B$ is the finite spherical entangling surface on the boundary, described by the inequality $\sum_i x_i^2 \le L^2$. The full interior of the causal diamond satisfies the inequality \cite{Verlinde:2019ade}
	\begin{equation}\label{eq:interior CD}
		L^2 - z^2 - \sum_i x_i^2 + x_0^2 \ge 2 L \abs{x_0}. 
	\end{equation}
	The transformation between the Poincar\'{e} and topological black hole metrics is given in Refs.~\cite{Casini:2011kv,Verlinde:2019ade}, which we do not repeat since the details are not important for the purpose of this paper. 
	
	\subsection{From Topological Black Hole to Kruskal–Szekeres Metric}
	\label{sub:KS metric}
	
	Our interest is in the dynamics of spacetime fluctuations near the light front of Rindler-AdS space. The light front coincides with the horizon of the topological black hole metric, where Eq.~\eqref{eq:metric_top_BH} becomes singular, and it becomes desirable to perform a coordinate transformation to overcome the apparent pathology of Eq.~\eqref{eq:metric_top_BH}. 
	We transform the topological black hole metric into the Eddington-Finkelstein (EF) metric as an intermediate step, defining the tortoise coordinate $\rho_{*}$
	\begin{equation} \label{eq:tortoise coord}
		\rho_{*} \equiv \int^\rho \frac{\dd{\rho'}}{f(\rho')} = \frac{L}{2} \ln \frac{\rho- L}{\rho + L},
	\end{equation}
	where $f(\rho) = \rho^2/L^2 -1$. Then we define two new coordinates $U$ and $V$
	\begin{equation}\label{eq:EF uv}
		V \equiv \tau + \rho_{*} \qand U \equiv \tau - \rho_{*}. 
	\end{equation}
	In terms of $U$ and $V$, the original topological black hole metric in Eq.~\eqref{eq:metric_top_BH} becomes
	\begin{align}
		d s^2 &= -f(\rho) dV^2 + 2 dV d\rho + \qty(\frac{\rho}{L})^2 d \Sigma_{d-2}^2, \qq{(EF-ingoing)}\label{eq:EF in} \\
		d s^2 &= -f(\rho) dU^2 - 2 dU d\rho + \qty(\frac{\rho}{L})^2 d \Sigma_{d-2}^2. \qq{(EF-outgoing)}\label{eq:EF out}
	\end{align}
	The metric $d \Sigma_{d-2}^2$ in the transverse space is given by
	\begin{equation} \label{eq:metric_hyperbolic_space}
		d \Sigma_{d-2}^2 = d \chi^2 + \sinh[2](\frac{\chi}{L}) d \Omega_{d-3}^2.
	\end{equation}
	The form of Eq.~\eqref{eq:metric_hyperbolic_space} plays an important role in determining the angular correlation functions of uncertainty in the photon traversal time. We will discuss angular correlations in detail in Sec.~\ref{sub:angular corr}. Both metrics above are non-singular at the horizon. While Eq.~\eqref{eq:EF in} describes the trajectories of particles on the upper half of the causal diamond in Fig.~\ref{fig:causal_diamond}, Eq.~\eqref{eq:EF out} describes the trajectories of particles on the lower half. 
	
	Following Refs.~\cite{Ahn:2019rnq,Misner:1973prb}, we define the ``light-cone'' coordinates in Rindler-AdS space
	\begin{equation} \label{coord: U  V}
		u = - L e^{-U/L} = - L \sqrt{\frac{\rho-L}{\rho+L}} e^{- \tau / L}, \qquad  v = L e^{V/L} =  L \sqrt{\frac{\rho-L}{\rho+L}} e^{ \tau / L},
	\end{equation}
	where the second equality relates $u$ and $v$ to the topological black hole coordinates $(\tau, \rho)$. Rindler-AdS space in the Kruskal-Szekeres metric becomes
	\begin{equation} \label{eq:metric Kruskal}
		d s^2 = - \frac{4 L^4 d u d v}{(L^2 + u v)^2} + \qty(\frac{L^2 - u v}{L^2 + u v})^2 d \Sigma_{d-2}^2.
	\end{equation}
	An advantage of the Kruskal–Szekeres metric is that the ``light-cone time'' $u$ and $v$ are proportional to the \textit{physical time} of a photon traveling inside an interferometer located on the (regularized) boundary at $z= z_c$. The proportionality constant turns out to be $(L / z_c)$ \cite{Verlinde:2019ade}, the conformal factor of the Poincar\'{e} metric. In summary, the elapsed lightcone time in traversing the lower (upper) causal diamond is $\D u = L$ ($\D v = L$), and the physical time $T_{\rm r.t.} \approx 2L^2 / z_c $.

		Our main task is to determine how spacetime fluctuations would alter the classical traversal time of the light beams; to do so, we start with studying metric perturbations about the Rindler-AdS background geometry in the subsequent section. 
		
		\section{Near Horizon Metric Perturbations} 
		\label{sec:NH_metric_fluc}
		
		Given the background geometry in Sec.~\ref{sec:background_geometry}, our goal is to study fluctuations on top of this background, and how they will give rise to a potentially observable effect in an interferometer experiment. Spacetime fluctuations are encapsulated by metric perturbations. Because these fluctuations are small in amplitude, we utilize the linearized Einstein equations to study the dynamics of the perturbed metric.
		
		The vacuum Einstein equations for AdS$_d$ spacetime reads
		\begin{equation} \label{eq:EInstein_tensor}
			G_{MN} \equiv R_{M N} - \frac{1}{2} g_{M N} R + \Lambda g_{M N} =0,
		\end{equation}
		where $M,N =1, \dots , d$ are the indices of AdS$_d$ bulk spacetime, and $\Lambda = - (d-1) (d-2 )/ 2 L^2 $ is the cosmological constant. All the metrics in Secs.~\ref{sec:introduction} and \ref{sec:background_geometry} are solutions to the vacuum Einstein equations. 
		
		We are interested in metric fluctuations in the near-horizon region of Rindler-AdS space, so it is most convenient to use the EF coordinates. Metric perturbations along the past (future) light front are described by Eq.~\eqref{eq:EF out} (Eq.~\eqref{eq:EF in}). We choose to study metric fluctuations along the past light front, which corresponds to using the EF-outgoing metric.  A completely analogous analysis applies for the future light front. The perturbed metric along the past light front is given by 
		\begin{align} \label{eq:pert EF out}
			\begin{split}
				d s^2 &= -f(\rho) dU^2 - 2 d U d\rho + \qty(\frac{\rho}{L})^2 d \Sigma_{d-2}^2 \\
				& + H_{UU} d U^2 + 2 H_{U \rho} d U d\rho + H_{\rho\rho} d \rho^2 + \cdots, 
			\end{split}
		\end{align}
		where $(\cdots)$ denotes $H_{a b}$, $a,b = 1, \dots, d-2$ in the transverse space. 
		
		The perturbed metric in Eq.~\eqref{eq:pert EF out} solves the linearized Einstein equation for $\abs{H_{M N}} \ll 1$, which in AdS space is given by \cite{Kovtun:2003wp,Kovtun:2005ev}
		\begin{equation} \label{eq:linear EE}
			\d G_{M N}^{(1)} \equiv R_{M N}^{(1)} + \frac{d-1}{L^2} H_{M N} = 0,
		\end{equation}
		where the perturbed Ricci tensor $R_{M N}^{(1)}$ satisfies \cite{Policastro:2002se}
		\begin{align} \label{eq:R_MN}
			R_{M N} = R_{M N}^{(0)} + R_{M N}^{(1)} + \cdots = -\frac{(d-1)}{L^2} (g^{(0)}_{M N} + H_{M N}).
		\end{align}
		Here, $g^{(0)}_{M N}$ denotes the background metric. Next, we expand the perturbations as a power series in the near-horizon region \cite{Ahn:2019rnq,Blake:2018leo}
		\begin{equation} \label{eq:series H_MN}
			H_{MN} = H_{M N}^{(0)} + H_{M N}^{(1)} \qty(\frac{r - L}{L}) + \cdots,
		\end{equation}
		and $H_{M N}^{(0)}$ can be written as 
		\begin{align} \label{eq:H_MN^(0)}
			H_{M N}^{(0)}(U, \vb{x}_\perp) = \int \frac{\dd{\o}}{2 \pi} \, h_{M N}(\vb{x}_\perp) e^{- i \o U},
		\end{align}
		where $\vb{x}_\perp$ denotes the coordinates in the transverse space, and $\omega$ is the frequency conjugate to $U$. Following the procedure in Refs.~\cite{Blake:2018leo,Ahn:2019rnq}, one can show that the $UU$-component of the linearized Einstein equation describes the Dray-'t Hooft shockwave perturbation in Refs.~\cite{DRAY1985173,Sfetsos:1994xa}. Substituting Eq.~\eqref{eq:H_MN^(0)} into Eq.~\eqref{eq:linear EE}, we find the $UU$-component of Eq.~\eqref{eq:linear EE} to be \cite{Ahn:2019rnq,Blake:2018leo}
		\begin{align} \label{eq:Guu full}
			\begin{split}
				&\frac{d-2}{L^2} \qty[1 + L\qty(4 \pi T - i \o -  \frac{3}{L})] h_{UU}^{(0)} - \laplacian_{\perp} h_{UU}^{(0)} \\
				&- \frac{i \o + 2 \pi T}{L} X = 0,
			\end{split}
		\end{align}
		where 
		\begin{equation} \label{eq:T hawking}
			T = \frac{f'(\rho)}{4 \pi} \bigg|_{\rho = L} = \frac{1}{2 \pi L}
		\end{equation}
		is the Hawking temperature. The variable $X$ denotes all $h_{MN}^{(0)}$ coupled to $h_{UU}^{(0)}$ via Eq.~\eqref{eq:linear EE}. In general, the exact form of $X$ is quite complicated. For instance, Ref.~\cite{Ahn:2019rnq} has computed the form of $X$ in AdS$_4$ to be
		\begin{equation} \label{eq:X 4d}
			X \stackrel{d = 4}{=} 2 \coth(\chi / L) h_{U \chi}^{(0)} + i \o L \qty(\csch[2](\chi / L) h_{\t\t}^{(0)} + h_{\chi\chi}^{(0)}) + 2 \csch[2](\chi / L) \pdv{h_{U \t}^{(0)}}{\t} + 2 L \pdv{h_{U \chi}^{(0)}}{\chi}.
		\end{equation}
		Fortunately, the precise form of $X$ will not be relevant for the purposes of this paper.
		
		Eq.~\eqref{eq:Guu full} thus imposes a constraint relating $h_{UU}^{(0)}$ to other metric perturbation components.  However, when $\o = \o_{\star} = i 2 \pi T$, the second line of Eq.~\eqref{eq:Guu full} vanishes altogether. The resulting equation takes on the same form of the partial differential equation describing metric perturbations due to gravitational shockwaves \cite{DRAY1985173,Sfetsos:1994xa}. As pointed out in Ref.~\cite{Blake:2018leo}, the point $\o_{\star} = i 2 \pi T$ is very special, as $2 \pi T $ is also known as the \textit{Lyapunov exponent}: $2 \pi T = \l_{\rm max}$, which characterizes chaotic behavior in a quantum system \cite{Shenker:2013pqa,Roberts:2016wdl,Shenker:2014cwa,Blake:2017ris}. Following the argument of Refs.~\cite{Blake:2018leo,Ahn:2019rnq}, one deduces that at the point $\o_{\star} = i 2 \pi T$, $h_{UU}^{(0)}$ decouples from the rest of $h_{MN}^{(0)}$ and becomes an \textit{independent} scalar degree of freedom which satisfies the equation 
		\begin{equation} \label{eq:shockwave PDE}
			\qty(\frac{d-2}{L^2} - \laplacian_{\perp}) h_{UU}^{(0)} = 0.
		\end{equation}
		The solution to this equation is readily obtained by setting $\laplacian_{\perp} h_{UU}^{(0)} = -k_\perp^2 h_{UU}^{(0)}$, with $k_\perp^2$ being the eigenvalue of the transversal Laplacian operator. Therefore, Eq.~\eqref{eq:shockwave PDE} is reduced to an algebraic equation 
		\begin{equation} \label{eq:algbraic relation}
			\frac{d-2}{L^2} + k_\perp^2 = 0.
		\end{equation}
		We can re-write Eq.~\eqref{eq:algbraic relation} by substituting $\o_{\star} = i 2 \pi T = L^{-1}$ into the expression
		\begin{equation}
			\o_{\star} =  i D k_\perp^2, \qquad D =  \frac{L}{d-2} = \frac{v_B}{2 \pi T},
		\end{equation}
		so it resembles the dispersion relation arising from a diffusive system. The diffusivity $D$ characterizes the so-called energy diffusion~\cite{Blake:2018leo}, because the metric perturbation is in the $UU$-component. The factor of $1/(d-2) = v_B$ has been shown~\cite{Ahn:2019rnq,Perlmutter:2016pkf} to be the \textit{butterfly velocity} in Rindler-AdS space. The butterfly velocity characterizes the speed of information propagating in a system with a horizon (\textit{e.g.,} a black hole), and it is closely related to the propagation of gravitational shockwaves and quantum chaos \cite{Shenker:2013pqa,Sekino:2008he,Susskind:2011ap,Shenker:2014cwa}. Furthermore, Ref.~\cite{Grozdanov:2017ajz} studying an AdS$_5$ black-brane obtained a similar diffusive dispersion with the same Lyapunov exponent, but with a different $v_B$. In fact, several recent works~\cite{Blake:2016sud,Blake:2016wvh,Blake:2017ris,Blake:2018leo} have shown that energy diffusion phenomenon is quite universal in various holographic systems, which all have the same Lyapunov exponent, but with a geometry-dependent $v_B$. 
		
		So far our discussions have been completely classical. Eq.~\eqref{eq:shockwave PDE} also describes classical gravitational shockwaves~\cite{DRAY1985173,Ahn:2019rnq,Blake:2018leo,Grozdanov:2017ajz} if we add a source, where the right-hand side is $8 \pi G_N T_{UU} \sim \ell_p^{d-2} \frac{\d(U - U_0)}{2\pi \tilde \ell_p} \delta^{d-2}({\bf x}_\perp - {\bf x}_\perp^0) $ for a classical shockwave stress-energy tensor propagating at $x_0 = (U_0,{\bf x}_\perp^0)$ with momentum $p_v = \frac{1}{2 \pi \tilde \ell_p}$. It is, however, possible to also consider quantum sources.  In particular, we focus on a quantum source from vacuum energy fluctuations, motivated by the 't Hooft commutation relations~\cite{tHooft:1996rdg,tHooft:2018fxg}.  In particular, Ref.~\cite{Verlinde:2022hhs} showed that vacuum fluctuations in Minkowski space, fixed by the 't Hooft commutation relations, give rise to the modular fluctuations in Eq.~\eqref{eq:DeltaK}. 
		In the following, we will utilize this result and apply it to Rindler-AdS space, by adding a quantum source to the vacuum Einstein equation Eq.~\eqref{eq:shockwave PDE} of size fixed by the 't Hooft commutation relations.  In so doing, we will reproduce quantum fluctuations in the round-trip photon travel time in Eq.~\eqref{eq:Uncert_T_rt_VZ2}.
		
		\subsection{Quantum Sources from the 't Hooft Commutation Relation}
		\label{sub:t'hooft comm}
		
		We will ultimately be interested in studying Eq.~\eqref{eq:shockwave PDE} in the presence of shockwaves from quantum fluctuations.  In particular, these quantum fluctuations are motivated by the commutation relations proposed in Ref.~\cite{Verlinde:2022hhs}, which are written in light-cone coordinates $(u, v)$. Thus, we will transform the EF coordinates $(U,\rho)$ of Eq.~\eqref{eq:shockwave PDE} to the light-cone  $(u,v)$ Kruskal–Szekeres metric, taking $h_{UU} \to h_{uu}$, where we suppress the superscript henceforth. Because $u$ and $U$ are related via Eq.~\eqref{coord: U  V}, it is straightforward to see that in the Kruskal–Szekeres metric, Eq.~\eqref{eq:shockwave PDE}, is 
		\begin{equation} \label{eq:PDE h_UU}
			\qty(\frac{d-2}{L^2} - \laplacian_{\perp}) h_{uu} = 0.
		\end{equation}
		Due to vacuum energy fluctuations, the right hand side of Eq.~\eqref{eq:PDE h_UU} is replaced with some stress-energy tensor $T_{uu}$. 
		
\begin{figure}
			\centering
			\hspace*{-0.5in}
			\includegraphics[width=0.5\textwidth]{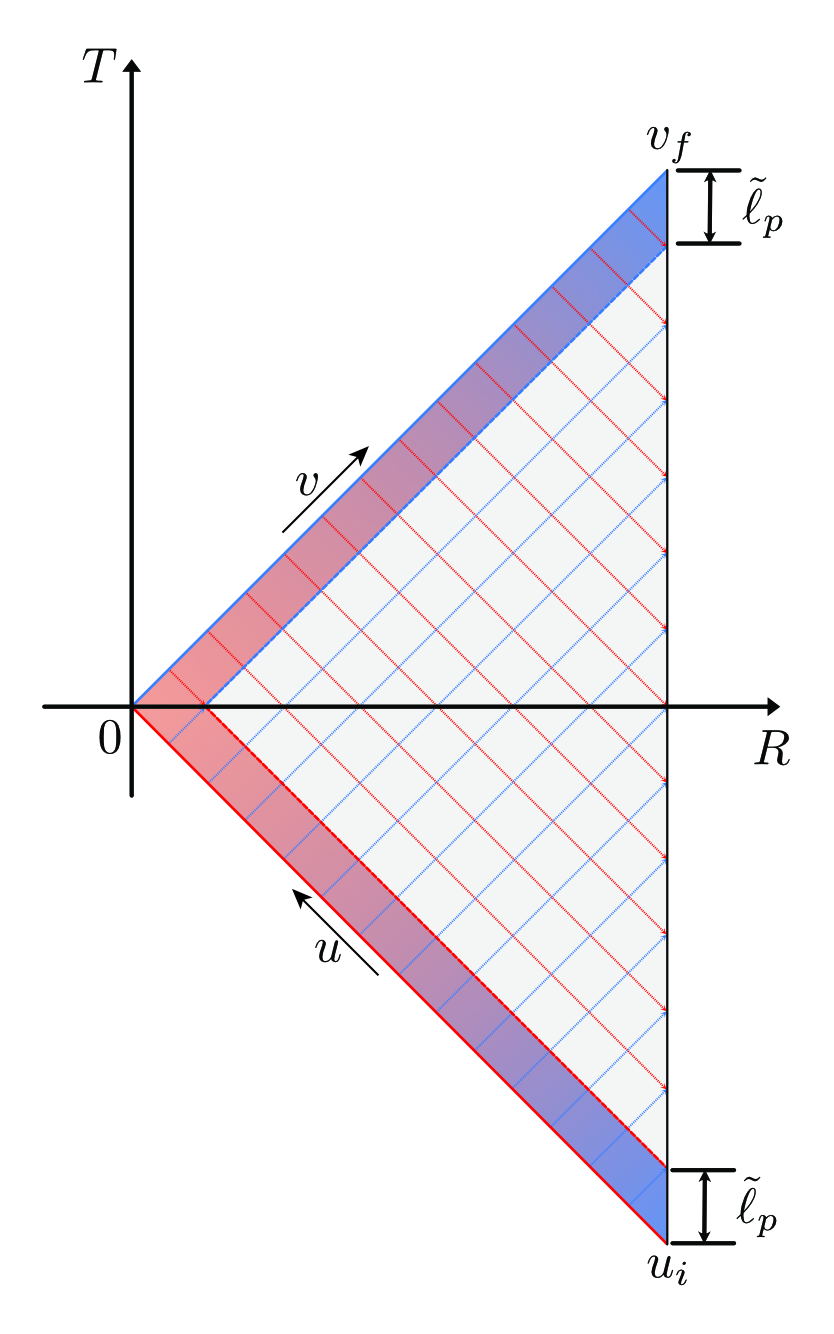} 
			\vspace*{-3mm}
			\caption{The causal diamond in Rindler-AdS space is foliated with a series nested causal diamonds. The separation between two adjacent diamonds is the decoherence length $\tilde{\ell}_p$. Each nested causal diamond intersects with the past (future) light front at a bifurcate horizon along the past (future) light front. The highlighted region corresponds to the near-light-sheet region of spacetime, where quantum fluctuations cause a probe photon to undergo random walk.}  \label{fig:nestedCD}
		\end{figure}

		Here we will assume that $T_{uu}$ captures the quantum nature of the fluctuations.  This is the {\em Ansatz} of this paper that differs from other literature, which further will be crucial for obtaining the fluctuation in the round-trip photon traversal time obtained in Ref.~\cite{Verlinde:2019ade}.
		In particular, we make use of a commutation relation (closely related to those proposed by 't Hooft) at \textit{unequal} times \cite{Verlinde:2022hhs}
		\begin{equation} \label{eq:comm T h}
			\comm{T_{uu}(x)}{h_{vv}(x')} = i \d^{d}(x - x'),
		\end{equation}
		where $x$ denotes the coordinates in Rindler-AdS$_d$ space, written in light-cone coordinates. The $d$-dimensional delta function can be factorized into three parts 
		\begin{equation} \label{eq:delta d-dim}
			\d^{d}(x - x') = \frac{1}{2}\d(u- u_0) \d(v - v_0) \d^{d-2}(\vb{x}_\perp - \vb{x}_\perp'),
		\end{equation}
		where $u_0$ and $v_0$ denote the location of the bifurcate horizon, and $\d^{d-2}(\vb{x}_\perp - \vb{x}_\perp')$ is the $(d-2)$-dimensional delta function in the transverse space. Note that the additional factor of $1/2$ comes from the normalization condition for the delta function in the Kruskal-Szekeres metric. Imposing the commutation relation in Eq.~\eqref{eq:comm T h} implies that $h_{uu}$ and $h_{vv}$ are no longer classical metric perturbations, but have been promoted to quantum operators. By further imposing the linearized Einstein equation on Eq.~\eqref{eq:comm T h}, we obtain an operator equation
		\begin{align} \label{eq:comm eq}
			\qty(\frac{d-2}{L^2} - \laplacian_{\perp}) \comm{h_{uu}(u, \vb{x}_\perp)}{h_{vv}(v, \vb{x}_\perp')} = \frac{i}{2} \ell_p^{d-2} \d (u - u_0) \d( v - v_0) \d^{d-2} (\vb{x}_\perp - \vb{x}_\perp').
		\end{align}
		Note that the transverse Laplacian acts only on $h_{uu}(u, \vb{x}_{\perp})$, and not the coordinates marked with a prime in $h_{vv}(v, \vb{x}_\perp')$. Eq.~\eqref{eq:comm eq} then implies that 
		\begin{equation} \label{eq:comm hUU hVV}
			\comm{h_{uu}(u, \vb{x}_\perp)}{h_{vv}(v, \vb{x}_\perp')} = \frac{i }{2}\ell_p^{d-2} \d(u - u_0) \d(v - v_0) f(\vb{x}_\perp; \vb{x}_\perp'),
		\end{equation}
		where $\ell_{p}^{d-2} = 8 \pi G_{N}^{(d)}$, and $f(\vb{x}_\perp; \vb{x}_\perp')$ is the Green function that satisfies 
		\begin{equation} \label{eq: Green G}
			\qty(\frac{d-2}{L^2} - \laplacian_{\perp}) f(\vb{x}_\perp; \vb{x}_\perp') = \d^{d-2}(\vb{x}_\perp - \vb{x}_\perp').
		\end{equation}
		Since $T_{uu}$ is a stochastic source in vacuum, this implies that $\expval{h_{uu}}$ and $\expval{h_{vv}}$ vanish, where $\expval{\cdots}$ denotes the expectation value of any minimum uncertainty state. However, the variance $\expval{h_{uu}^2}$ and $\expval{h_{vv}^2}$ are non-vanishing by the virtue of the Robertson uncertainty relation in quantum mechanics
		\begin{equation} \label{eq:Robertson uneq}
			\expval{h_{uu}^2} \expval{h_{vv}^{2}} = \abs{\frac{1}{2 i} \expval{\comm{h_{uu}}{h_{vv}}}}^2 = \qty(\frac{\ell_p^{d-2}}{4})^{2} \qty[\d(u - u_0) \d(v- v_0) f(\vb{x}_\perp; \vb{x}_\perp')]^2.
		\end{equation} 
		Two important comments are in order.
		\begin{enumerate}[(1)]
			\item Formally, the 't Hooft commutation relations were formulated on the horizon of a black hole. In the present context, that would imply Eq.~\eqref{eq:Robertson uneq} is evaluated at the bifurcate horizon, which is located at $u_0 = v_0 = 0$. However, according to Refs.~\cite{Banks:2021jwj,Zurek:2022xzl}, the light beam in an interferometer system passes through a series of causal diamonds. Specifically, the maximal causal diamond in Fig.~\ref{fig:causal_diamond} is foliated by a sequence of nested causal diamonds. Each of the adjacent causal diamonds is separated by a length scale, called the \textit{decoherence length} $\tilde{\ell}_p$, given in Eq.~\eqref{eq:ell_p tilde ansatz}. Subsequent causal diamonds separated by a distance larger than $\tilde{\ell}_p$ become \textit{statistically independent} \cite{Banks:2021jwj}.  A schematic of the nested causal diamonds is shown in Fig.~\ref{fig:nestedCD}. Along the past light front, we keep $v$ fixed, while $u$ varies along the trajectory; along the future light front, the opposite holds.  From the viewpoint of nested causal diamonds, a light beam traveling along the past light front will experience a series of statistically independent fluctuations. Along the past (future) null trajectory, where the clock is $u~(v)$, one can define the variance $\expval{h_{uu}^2} \equiv \expval{h_{uu}(u, \vb{x}_\perp) h_{uu}(u', \vb{x}_\perp')}$ ($\expval{h_{vv}^2} \equiv \expval{h_{vv}(v, \vb{x}_\perp) h_{vv}(v', \vb{x}_\perp')}$), where $u'~(v')$ denotes the location of the bifurcate horizons of each nested causal diamond. 
			\item We postulate the past (future) light front will be smeared out by $\tilde \ell_p$, which operationally means the delta function which localizes the light front at $v_0~(u_0)$, $\d(v - v_0 ) =\d (v)$ ($\d(u - u_0 ) =\d (u)$) will be replaced by a regularized delta function of Planckian width. This is quite similar to the implementation of a stretched horizon for a black hole in Refs.~\cite{Susskind:1993if,Sekino:2008he,Susskind:2011ap}. In the present case, we implement the ``smearing'' of the light front by regularizing $\d(v)$ with a Poisson kernel of a Lorentzian width $\tilde{\ell}_p$:
			\begin{equation} \label{eq:lorentzian}
				\d(v) = \lim_{\tilde{\ell}_p \to 0} \frac{2}{\pi} \frac{\tilde{\ell}_p}{\tilde{\ell}_p^2 + v^2} \approx\frac{2}{\pi \tilde{\ell}_p} \qq{along the past light front $v \to 0$.}
			\end{equation}
		\end{enumerate}
		These points are illustrated in Fig.~\ref{fig:nestedCD}, where the broadening of the delta function along the past and future light front is shown as a shaded red/blue gradient.  Note that our final result for the fluctuations in the photon round-trip traversal time may depend on the precise form of the delta-function regularization by an ${\cal O}(1)$ number, but can be absorbed into an ${\cal O}(1)$ (dimensionless) coefficient by matching the present hydrodynamic result to the earlier result in Eq.~\eqref{eq:Uncert_T_rt_VZ2}.  The regularization scheme thus will not impact the overall physical picture since the dimensionful scales match between the present hydrodynamic calculation and the result of Ref.~\cite{Verlinde:2019ade}.
		
		In summary, Eq.~\eqref{eq:Robertson uneq} together with Eq.~\eqref{eq:lorentzian}, at a {\em fixed point} on the past or future null horizon, implies a non-vanishing two-point function of $h_{uu}$ and $h_{vv}$ given by
		\begin{align}
			\expval{h_{uu}(u, \vb{x}_\perp) h_{uu}(u', \vb{x}_\perp')} &= \frac{\ell_p^{d-2}}{2\pi \tilde{\ell}_p} \d(u-u') f(\vb{x}_\perp;\vb{x}_\perp'), \label{eq:two-point Huu}\\
			\expval{h_{vv}(v, \vb{x}_\perp) h_{vv}(v', \vb{x}_\perp')} &= \frac{\ell_p^{d-2}}{2\pi \tilde{\ell}_p} \d(v-v') f(\vb{x}_\perp;\vb{x}_\perp'). \label{eq:two-point Hvv}
		\end{align}
		In the next Section, we study how these fundamental commutators can be evolved to give the integrated uncertainty in the light traversal time.
		
		\section{Uncertainty in Photon Traversal Time From Near-horizon Quantum Fluctuations} 
		\label{sec:quant_noise}
		
		In the previous section, we have argued that vacuum energy fluctuations in the near-horizon region give rise to non-vanishing variance of the metric perturbations, Eqs.~\eqref{eq:two-point Huu} and \eqref{eq:two-point Hvv}. 
		The equations that govern the two-point function of $h_{uu}$ and $h_{vv}$ are shown in Eq.~\eqref{eq:two-pt hh}.
		Note that the two-point functions of metric perturbations themselves are not observables in an interferometer system. To connect the equations above to a quantity more directly connected to the observable, we first define two ``light-ray'' operators as in Ref.~\cite{Verlinde:2022hhs}:
		\begin{align}
			X^v &= v + \int^{u} \dd{u'} h_{uu}(u', \vb{x}_\perp), \label{eq:Xv}\\ 
			X^u &=  u + \int^v \dd{v'} h_{vv}(v', \vb{x}_\perp). \label{eq:Xu}
		\end{align}
		Ref.~\cite{Verlinde:2022hhs} has demonstrated that the 't Hooft commutation relations in Eq.~\eqref{eq:comm T h} applied on the bifurcate horizon implies a commutation relation of $X^u$ and $X^v$:
		\begin{equation} \label{eq:comm X X}
			\comm{X^u(\vb{x}_{\perp})}{X^v(\vb{x}_{\perp}')} = i \ell_{p}^{d-2} f(\vb{x}_{\perp}; \vb{x}'_{\perp}).
		\end{equation} 
		Presently, we are interested in obtaining the accumulated uncertainty along the light sheet. This uncertainty is computed from solving the following equations:
		\begin{align}
			\qty(\frac{d-2}{L^2} - \laplacian_{\perp}) \expval{\pd_{u}X^v(u,\vb{x}_\perp) \pd_{u'}X^v(u',\vb{x}_\perp') } &= \frac{\ell_p^{d-2}}{2 \pi \tilde{\ell}_{p}} \d(u - u') \d^{d-2}(\vb{x}_\perp - \vb{x}_\perp'), \label{eq:langevin Xv} \\
			\qty(\frac{d-2}{L^2} - \laplacian_{\perp}) \expval{\pd_{v}X^u(v,\vb{x}_\perp) \pd_{v'}X^u(v',\vb{x}_\perp') } &= \frac{\ell_p^{d-2}}{2 \pi \tilde{\ell}_{p}} \d(v - v') \d^{d-2}(\vb{x}_\perp - \vb{x}_\perp'). \label{eq:langevin Xu} 
		\end{align}
		Eqs.~\eqref{eq:langevin Xv} and \eqref{eq:langevin Xu} look similar to a Langevin equation that describes random motion of a particle suspending in a dissipative fluid~\cite{langevin1908}. Recall that in statistical mechanics, the one-dimensional Langevin equation is a stochastic differential equation which takes on the form (for a massless particle) of Eq.~\eqref{eq:langevin}.
		Besides the spatial response in the transverse plane, we can clearly identify quantities derived from gravitational shockwave dynamics in Eqs.~\eqref{eq:langevin Xv} and \eqref{eq:langevin Xu} with the dynamics of a microscopic particle in a fluid subjected to a random force. In particular, we find the following identifications: 
		\begin{align} \label{eq:gravity langevin id}
			\begin{split}
				\dot{X}(\tau,\vb{x}_\perp) &=  
				\begin{cases}
					\pd_{u} X^v(u, \vb{x}_\perp) \qq{past light front,}\\
					\pd_{v} X^u(v, \vb{x}_\perp) \qq{future light front.}
				\end{cases}\\
				\expval{F(\tau,\vb{x}_{\perp}) F(\tau', \vb{x}'_{\perp})} &\sim
				\begin{cases}
					\d(u - u') f(\vb{x}_{\perp} - \vb{x}'_{\perp}) \qq{past light front,}\\
					\d(v- v') f(\vb{x}_{\perp} - \vb{x}'_{\perp}) \qq{future light front.}
				\end{cases}
			\end{split}
		\end{align}
		We now compute the observable from Eqs.~\eqref{eq:langevin Xv} and \eqref{eq:langevin Xu}. An appropriate observable in an interferometer is the total time delay of a light beam traversing the whole causal diamond. This time-delay is measured on the boundary in the Poincar\'{e} metric: $T_{\rm r.t.} = L x_0/ z_c$. To compute the total time delay, we need to integrate over all {\em local} (and statistically uncorrelated at distinct spacetime points) fluctuations generated by the quantum uncertainty in Eqs.~\eqref{eq:langevin Xv} and \eqref{eq:langevin Xu}: 
		\begin{align} 
			\qty(\frac{d-2}{L^2} - \laplacian_{\perp})\expval{X^v(\vb{x}_\perp) X^v(\vb{x}_\perp')} &= \frac{\ell_{p}^{d-2} \d^{d-2}(\vb{x}_{\perp} - \vb{x}_{\perp}')}{2 \pi \tilde{\ell}_p} \int_{u_i}^{0} \dd{u} \int_{u_i}^{0} \dd{u'} \d(u-u') \nonumber  \\
			&= -\frac{\ell_{p}^{d-2}}{2 \pi \tilde{\ell}_p} u_i \d^{d-2}(\vb{x}_{\perp} - \vb{x}_{\perp}'), \label{eq:XvXv} \\
			\qty(\frac{d-2}{L^2} - \laplacian_{\perp}) \expval{X^u(\vb{x}_\perp) X^u(\vb{x}_\perp')} &= \frac{\ell_{p}^{d-2} \d^{d-2}(\vb{x}_{\perp} - \vb{x}_{\perp}')}{2 \pi \tilde{\ell}_p} \int_{0}^{v_f} \dd{v} \int_{0}^{v_f} \dd{v'} \d(v-v')  \nonumber \\
			&= \frac{\ell_{p}^{d-2}}{2 \pi \tilde{\ell}_p} v_f \d^{d-2}(\vb{x}_{\perp} - \vb{x}_{\perp}'). \label{eq:XuXu}
		\end{align}
		The integration limits in Eqs.~\eqref{eq:XvXv} and \eqref{eq:XuXu} are:
		\begin{equation}
			v_f = -u_i = L. 
		\end{equation}
		Notice that these equations already exhibit the random-walk behavior,  shown in Eq.~\eqref{eq:var Trt ansatz}, proposed in Refs.~\cite{Verlinde:2019xfb,Banks:2021jwj,Zurek:2022xzl}, where the total uncertainty in a length operator accumulates linearly with the size of the causal diamond (which here is given by $u_i,~v_f$).
		
		We first consider the case in which the full symmetry of the transverse space is respected (corresponding to the $s$-wave mode), in order to directly compare with Eq.~\eqref{eq:Uncert_T_rt_VZ2}. To extract this information from our analysis (which includes the transverse response), we thus {\em (i)} take the operator $\laplacian_{\perp} \rightarrow 0$ in the left-hand side of Eqs.~\eqref{eq:XvXv},~\eqref{eq:XuXu}, and {\em (ii)} integrate $\d^{d-2}(\vb{x}_{\perp}-\vb{x}'_{\perp})$ over the area and then divide by the area of $\Sigma_{d-2}$ in the right-hand side of Eqs.~\eqref{eq:XvXv},~\eqref{eq:XuXu}:
		\begin{equation} \label{eq:symmetric}
			(i):\frac{d-2}{L^2} - \laplacian_{\perp}\to \frac{d-2}{L^2}, \qquad  
			(ii):\d^{d-2}(\vb{x}-\vb{x}_{\perp}') \to \frac{1}{A(\Sigma_{d-2})}\int \d^{d-2}(\vb{x}-\vb{x}') \dd{\Sigma_{d-2}}=\frac{1}{A(\Sigma_{d-2})} .
		\end{equation}

		In an interferometer, the quantity related to the observable is the round-trip time of a light beam measured by a clock on the boundary located at $z = z_c$~\cite{Verlinde:2019ade}: 
		\begin{equation} \label{eq:Vf-Ui}
			T_{\rm r.t.}= \frac{L}{z_c} (v_f - u_i) \approx \frac{L}{z_c} (2 L).
		\end{equation}
		Fluctuations of $T_{\rm r.t.}$ are captured by the two-point function in Eqs.~\eqref{eq:XvXv} and \eqref{eq:XuXu}, which is now found to be:
		\begin{equation} \label{eq:symmetric rel uncert}
			\begin{split}
				\frac{\D T_{\rm r.t.}^2}{T_{\rm r.t.}^2} &\equiv \frac{1}{T_{\rm r.t.}^2} \qty(\expval{X^v(\vb{x}_\perp) X^v(\vb{x}_\perp')} + \expval{X^u(\vb{x}_\perp) X^u(\vb{x}_\perp')} ) \\
				&= \frac{1}{2 (d-2)}\qty( \frac{L}{\tilde{\ell}_p} )\frac{1}{S_{\rm ent.}}.
			\end{split}
		\end{equation}
		Here we have used the definition of the entanglement entropy $S_{\rm ent.} \equiv A(\Sigma_{d-2}) / 4 G_N^{(d)}$ with $ 8 \pi G_{N}^{(d)} = \ell_{p}^{d-2} $. 
		Comparing our result in Eq.~\eqref{eq:symmetric rel uncert} with that from Ref.~\cite{Verlinde:2019ade} (shown in Eq.~\eqref{eq:Uncert_T_rt_VZ2}) allows us to determine $\tilde{\ell}_p$ as
		\begin{equation} \label{eq:tilde_ell_p}
			\tilde{\ell}_p = \frac{L}{4\sqrt{S_{\rm ent.}}}.
		\end{equation}
	This is the same length scale first identified, through independent and complementary means, in Refs.~\cite{Verlinde:2022hhs,Banks:2021jwj,Zurek:2022xzl} and quoted in Eq.~\eqref{eq:ell_p tilde ansatz}.  In particular, Ref.~\cite{Verlinde:2022hhs} identified $\tilde \ell_p$ as the relevant uncertainty scale appearing in the commutation relation Eq.~\eqref{eq:comm X X}, giving a physical interpretation to the width of the stretched horizon we have employed here.  The dependence of the uncertainty scale on the dimensionful scales $\ell_p,~L$ can be parametrically seen by noting that $f({\bf x}_\perp,{\bf x}'_\perp) \sim L^{4-d}$ (as can be seen from Eq.~\eqref{eq: Green G} and which we will write out explicitly below), such that that right-hand side of the uncertainty relation Eq.~\eqref{eq:comm X X} has a dimensionful scaling as $\ell_p^{d-2}/L^{d-4} \sim \tilde \ell_p^2$.  Even more precisely, Eq.~\eqref{eq:tilde_ell_p} agrees to a factor of 4 (which can be attributed to uncertainty due to the regularization procedure employed here) with that predicted in Refs.~\cite{Banks:2021jwj,Zurek:2022xzl}.
		
		\subsection{Angular Correlations of Photon Traversal Time Fluctuations}
		\label{sub:angular corr}
		
		The form of the expressions in Eqs.~\eqref{eq:XvXv} and \eqref{eq:XuXu} allows us to now also extract the angular correlations, via $f(\vb{x}_{\perp}; \vb{x}'_{\perp})$, which as the Green function of the transversal Laplacian in~\eqref{eq: Green G} becomes
		\begin{equation} \label{eq:helmholtz equation}
			\qty[\frac{d-2}{L^2}-\pdv[2]{}{\chi} - (d-3) \coth(\frac{\chi}{L})\frac{1}{L}\pdv{}{\chi} - \frac{1}{L^2 \sinh[2](\chi / L)} \laplacian_{\mathbf{S}^{d-3}} ] f(\vb{x}_\perp; \vb{x}_\perp') = \d^{d-2}(\vb{x}_\perp - \vb{x}_\perp'),
		\end{equation}
		where the Laplacian operator on the transverse space $\Sigma_{d-2} \cong \mathbf{H}^{d-2}$ is given by Ref.~\cite{Fundamental}  and $\laplacian_{\mathbf{S}^{d-3}}$ denotes the Laplacian on a $(d-3)$-dimensional unit-sphere. In the following, we consider an interferometer setup in which the two end mirrors are located at $\chi$ and $\chi'$. In other words, the two interferometer arms pick out two particular directions in the $\chi$-coordinate of the transverse space, while leaving the residual subspace $\mathbf{S}^{d-3}$ invariant. Therefore, we can neglect the term $\laplacian_{\mathbf{S}^{d-3}}$ in Eq.~\eqref{eq:helmholtz equation}. Spherical symmetry implies that the solution of Eq.~\eqref{eq:helmholtz equation} depends only on the geodesic distance in $\Sigma_{d-2}$, which is given by \cite{Ahn:2019rnq,Fundamental}
		\begin{equation} \label{eq: geo dist}
			\xi(\vb{x}_\perp; \vb{x}_\perp') \equiv \cosh[-1]( \cosh(\frac{\chi}{L}) \cosh(\frac{\chi'}{L}) - \sinh(\frac{\chi}{L}) \sinh(\frac{\chi'}{L}) \cos \gamma),
		\end{equation}
		where $\gamma$ is the polar angle subtended by the two interferometer arms. To further simplify the problem we consider the case in which $L$ is sufficiently large compared to $\xi$ such that the term $(d-2)/L^2$ can be neglected. Eq.~\eqref{eq:helmholtz equation} then reduces to
		\begin{equation} \label{eq:laplace equation}
			- \pdv[2]{f(\vb{x}_\perp;\vb{x}_\perp')}{\chi} - \frac{(d-3)}{L} \coth(\frac{\chi}{L}) \pdv{f(\vb{x}_\perp;\vb{x}_\perp')}{\chi} = \d^{d-2}(\vb{x}_\perp-\vb{x}_\perp').
		\end{equation}
		The solution to Eq.~\eqref{eq:laplace equation} is found in Ref.~\cite{Fundamental} to be 
		\begin{equation} \label{eq:sol f(x,x')}
			f(\vb{x}_\perp; \vb{x}_\perp') = \frac{1}{ \Omega_{d-3} L^{d-4}} f(\Sigma; \Sigma').
		\end{equation}
		where $f(\Sigma; \Sigma')$ is given in terms of the hypergeometric function \cite{Fundamental,Ahn:2019rnq}
		\begin{equation} \label{eq:hypergeom func}
			f(\Sigma;\Sigma') = \frac{1}{(d-3) \cosh^{d-3} \xi}{}_{2}F_{1} \qty(\frac{d-3}{2}, \frac{d-2}{2}; \frac{d-1}{2}; \frac{1}{\cosh^2\xi}).
		\end{equation}
		In the limit where the hyperboloid $\mathbf{H}^{d-2}$ looks locally Euclidean, $f(\Sigma;\Sigma')$ reduces to the familiar result in Euclidean space \cite{Fundamental}:
		\begin{equation} \label{eq:f(Sigma;Sigma')}
			f(\Sigma; \Sigma')
			\approx 
			\begin{cases}
				\ln \frac{L}{\abs{\chi-\chi'}}&\qfor d = 4, \\
				\qty(\frac{L}{\abs{\chi - \chi'}})^{4-d}&\qfor d \ge 5.
			\end{cases}
		\end{equation}
		Because the Green function $f(\vb{x}_{\perp};\vb{x}'_{\perp}) \sim L^{4-d}$, it receives a conformal rescaling on the (regularized) boundary at $z = z_c$ in the Poincar\'{e} metric
		\begin{equation} \label{eq:f(x,x') bdy}
			f(\vb{x}_{\perp}; \vb{x}'_{\perp}) \stackrel{z = z_c}{\to} \qty(\frac{L}{z_c})^{4-d} f(\vb{x}_{\perp}; \vb{x}'_{\perp}).
		\end{equation}
		
		Accounting for the conformal factor $(L/z_c)$ properly, and using Eq.~\eqref{eq:XvXv} together with~\eqref{eq:XuXu}, we find the fluctuations $\D T_{\rm r.t.}^2$ to be
		\begin{align} \label{eq:var Trt}
			\begin{split}
				\D T_{\rm r.t.}^2(\vb{x}_{\perp};\vb{x}'_{\perp})&\equiv \left(\frac{L}{z_c}\right)^2 \left(\expval{X^v(\vb{x}_\perp) X^v(\vb{x}_\perp')} + \expval{X^u(\vb{x}_\perp) X^u(\vb{x}_\perp')} \right)\\
				&=\qty(\frac{L}{z_c})^2 \qty(\frac{2L^3}{\tilde{\ell}_p})  \qty[\frac{\ell_p^{d-2}}{ 2 \pi\Omega_{d-3}L^{d-2}(L / z_c)^{d-3}}] f(\Sigma; \Sigma')\\
				&= \qty(\frac{L}{z_c})^2 8 L^{2} \frac{1}{\sqrt{S_{\rm ent.}}} f(\Sigma; \Sigma').
			\end{split}
		\end{align}
		In evaluating the second line, we substituted the area of the transverse space $A(\Sigma_{d-2}) \approx\Omega_{d-3} L^{d-2} \qty(L/z_c)^{d-3}$ (for $z_c \to 0$) and used the definition of the entanglement entropy again, while in the third line we have used the scale precisely identified in Eq.~\eqref{eq:tilde_ell_p}.
		Thus, the relative uncertainty of photon round trip time, with angular correlations, is: 
		\begin{equation} \label{eq:rel uncert}
			\frac{\D T_{\rm r.t.}^2}{T_{\rm r.t.}^2}(\vb{x}_{\perp};\vb{x}'_{\perp}) = \frac{2}{\sqrt{S_{\rm ent.}}} f(\Sigma; \Sigma'). 
		\end{equation}

		\section{Summary and Discussion} \label{sec:summery} 
		
		In this paper, we have shown that vacuum energy fluctuations in AdS space, with a quantum noise term motivated by commutation relations presented in Ref.~\cite{Verlinde:2022hhs} and shown in Eqs.~\eqref{eq:comm T h} and~\eqref{eq:comm X X}, give rise to hydrodynamic behavior for the fluctuations of the spacetime geometry. In particular, we demonstrated that the near-horizon fluctuations of a finite causal diamond is a diffusive process that captures ``random walk'' characteristics in time (but with transverse spatial coorelations) of quantum spacetime fluctuations. We further analyzed the effect of these fluctuations on the traversal time of photons traveling from the boundary and reflecting off a mirror in the bulk, confirming the previous result of Ref.~\cite{Verlinde:2019ade} despite taking a computationally complementary route. An important step in our reasoning was to focus only on the hydrodynamics on the stretched horizon of a causal diamond, distinct from the usual fluid/gravity correspondence that proposes a duality between the bulk gravitational perturbations and boundary hydrodynamics.   
		
		There are many interesting future directions to pursue. First, one could carry out a similar type of analysis in Minkowski space.  
		Second, one could seek to understand the underlying origins of these vacuum fluctuations from shockwave geometries. Finally, one could utilize theoretical tools such as out-of-time-order correlators (OTOCs) that describe fast-scrambling systems and quantum chaos to study the connection between hydrodynamics and shockwave geometries. We look forward to further developments in these formal aspects and its groundwork for future observational tests.

		\section*{Acknowledgments}
		We thank Tom Banks, Temple He, Cynthia Keeler, Vincent Lee, Allic Sivaramakrishnan and Erik Verlinde for discussion on these directions. We are supported by the Heising-Simons Foundation “Observational Signatures of Quantum Gravity” collaboration grant 2021-2817. The work of KZ is also supported by a Simons Investigator award and the U.S. Department of Energy, Office of Science, Office of High Energy Physics, under Award No. DE-SC0011632.

		
		\bibliography{bibliography}
		
	\end{document}